\documentclass[12pt, reqno]{amsart}
\usepackage{amsmath, amsthm, amscd, amsfonts, amssymb, graphicx, color}
\usepackage[bookmarksnumbered, colorlinks, plainpages]{hyperref}

\usepackage[table]{xcolor}
\usepackage{times}
\usepackage{amssymb, amsmath}
\usepackage{amsthm}
\usepackage{tikz}
\usepackage{bm}
\usetikzlibrary{matrix,arrows}
\usepackage{enumitem}
\usepackage{subcaption}
\usepackage{booktabs}
\usepackage[section]{placeins}
\usepackage{float}

\theoremstyle{plain}

\theoremstyle{remark}

\definecolor{Gray}{gray}{0.95}
\newcolumntype{g}{>{\columncolor{Gray}}c}

\graphicspath{{./images/}}

\begin{document}
\setcounter{page}{1}

\centerline{}

\centerline{}

\title[New Era of Artificial Intelligence in Education]{New Era of Artificial Intelligence in Education: Towards a Sustainable Multifaceted Revolution}

\author[F. Kamalov.]{Firuz Kamalov$^1$$^{*}$}

\address{$^{1}$Faculty of Engineering, Canadian University Dubai, Dubai, UAE.}
\email{\textcolor[rgb]{0.00,0.00,0.84}{firuz@cud.ac.ae}}

\author[D. Santandreu Calonge]{David Santandreu Calonge$^2$}

\address{$^{2}$Academic Development, Mohamed bin Zayed University of Artificial Intelligence, Abu Dhabi, UAE.}
\email{\textcolor[rgb]{0.00,0.00,0.84}{david.santandreu@mbzuai.ac.ae}}

\author[I. Gurrib.]{Ikhlaas Gurrib$^3$}

\address{$^{3}$Faculty of Management, Canadian University Dubai, Dubai, UAE.}
\email{\textcolor[rgb]{0.00,0.00,0.84}{ikhlaas@cud.ac.ae}}



\keywords{AI, deep learning, education, intelligent systems, ChatGPT}

\date{\today
\newline \indent $^{*}$ Corresponding author}

\begin{abstract}
The recent high performance of ChatGPT on several standardized academic tests has thrust the topic of artificial intelligence (AI) into the mainstream conversation about the future of education. As deep learning is poised to shift the teaching paradigm, it is essential to have a clear understanding of its effects on the current education system to ensure sustainable development and deployment of AI-driven technologies at schools and universities. 
This research aims to investigate the potential impact of AI on education through review and analysis of the existing literature across three major axes: applications, advantages, and challenges.
Our review focuses on the use of artificial intelligence in collaborative teacher--student learning, intelligent tutoring systems, automated assessment, and personalized learning. We also report on the potential negative aspects, ethical issues, and possible future routes for AI implementation in education. Ultimately, we find that the only way forward is to embrace the new technology, while implementing guardrails to prevent its abuse.
\end{abstract} \maketitle

\section{Introduction}
Artificial intelligence (AI) has quickly established itself as a 
transformative force in a wide range of industries, including education. The development of AI has resulted in an array of advancements and innovations that have impacted many facets of human life. As a fundamental component to societal evolution and individual development, education has had significant benefits from AI breakthroughs.  The integration of AI in educational systems is altering the ways in which students learn, teachers educate, and institutions function. By personalizing learning experiences, automating administrative responsibilities, and delivering real-time feedback, AI is revolutionizing the educational landscape, bridging gaps, and encouraging a more inclusive and effective learning environment. {Given the importance of integrating AI in education, there is a need to reflect on its implications.}

{Our goal is to study the potential impact of AI on education based on the review of the current literature. We focus on three major themes: applications, advantages, and challenges. The review procedure consists of searching the Scopus database using the relevant keywords. The results are filtered, sorted, and analyzed to extract the pertinent information. We identify several subfields within each theme that are used to categorize the literature and structure the problem in a more coherent fashion.
The results are presented in several sections according to the major themes and subthemes. We find that the best way forward is to embrace the new technology, while implementing guardrails to prevent \mbox{its abuse}.}

One of the key applications of AI is natural language processing (NLP). The aim of NLP is to develop intelligent systems that can understand human text and speech. In particular, intelligent chatbots have been increasingly deployed in various industries to provide customer service and support other tasks \cite{Nicolescu}. The development of modern chatbots began in 2016 and has accelerated up to the current date \cite{Adamopoulou}. The advent of chatbots has also affected the field of education \cite{Hwang2021}. A recent survey found that the use of chatbots in education has been steadily increasing \cite{Wollny}. Several studies found that chatbots can improve students' learning experiences and facilitate their education \cite{Okonkwo}.

\begin{figure}[h]

\includegraphics[trim={0cm 0cm 2.5cm 0.1cm},clip, width=0.9\textwidth]{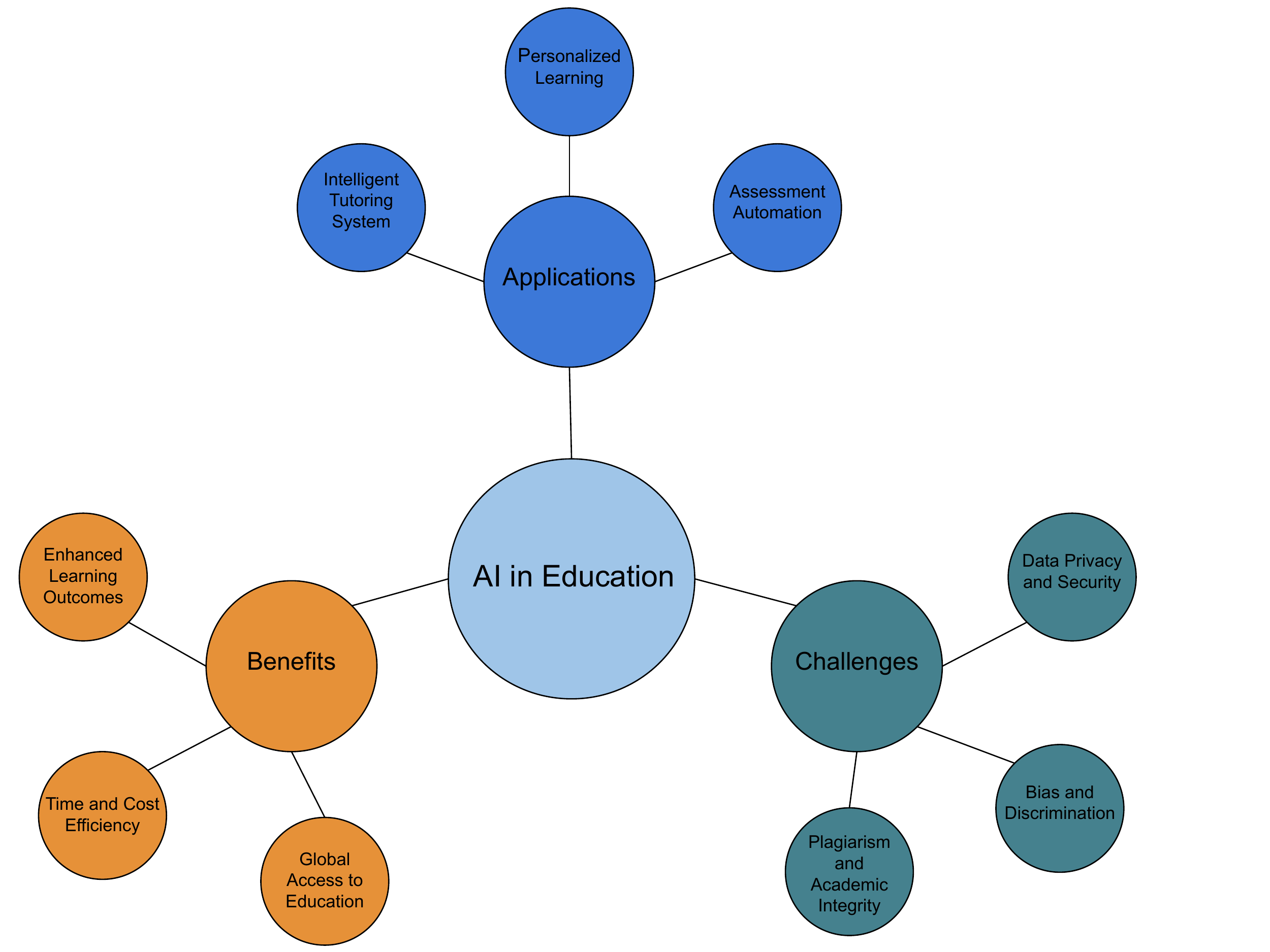}
\caption{Multifaceted impact of AI in education.}
\label{ai_chart}
\end{figure}

The turning point in the adoption of AI in society came in November, 2022 with the release of ChatGPT. The advanced writing and comprehension abilities of ChatGPT surprised many people, earning a wide-ranging audience and garnering unprecedented {attention}. 
 It was the first time that an audience outside the machine learning community truly realized the potential and immediacy of AI. The education sector was arguably the most affected by ChatGPT. The potential of ChatGPT to deliver intelligent tutoring systems, on one hand, and as a tool for academic dishonesty, on the other hand, has sparked an intense debate. Educators at secondary and tertiary education institutions have raised the alarm over the possibility of the abuse of ChatGPT by students and called for its restrictions. School districts in Australia’s Queensland and Tasmania schools and New York City and Seattle have prohibited the use of ChatGPT on students’ devices and networks. Several universities, colleges, and schools are evaluating similar restrictions \cite{Wilcox}. 
However, it appears impossible to prevent the students from using AI. As highlighted in  \cite{Stokel}, ChatGPT has great potential to provide solutions to college students on a range of tasks from essay writing to code creation.  Ultimately, the best way forward is to incorporate AI into the educational system and leverage its capabilities to deliver better learning outcomes for the students. In order to advance the debate over the optimal approach to utilizing AI, in this paper, we investigate its potential benefits as well dangers (Figure \ref{ai_chart}).

The potential applications of AI in education include personalized learning, intelligent tutoring systems, automation of assessment, and teacher--student collaboration \cite{Wollny}. Personalized learning is possible given the scalability of AI to the entire student population. AI algorithms such as reinforcement learning can be used to dynamically learn about the individual needs of a student and adapt the learning process accordingly. In connection with personalized learning, intelligent tutoring systems can be developed that can actively interact with students, giving valuable feedback. Another impactful aspect of AI is automation of the assessments. Computer vision and natural language processing systems can be combined to automatically grade homework, quizzes, and exams. Automated grading will provide a tremendous relief to instructors, giving them more time to spend with students. AI can also be useful in facilitating teacher--student collaboration by providing various feedback and analytics.

The applications of AI in education highlight the potential for huge advantages that are made possible by intelligent systems. The impact of AI can be seen in improved learning outcomes, time and cost efficiency, global access to quality education, and other benefits. Personalized learning and intelligent tutoring systems can help improve learning outcomes for students, especially in underserved populations. The global reach and scalability of AI will allow students from both developed as well as developing nations to benefit from better learning experiences. Automated grading will have massive cost- and time-saving benefits in education. Currently, around 40\% of teachers' time is spent on grading and related activities. Without the burden of grading, teachers will be able to spend more time with students and provide more learning support.

While the applications and benefits of AI in education can paint an alluring picture, it is important to be aware of potential hazards of introducing autonomous systems in education. Since children are more susceptible than adults to misinformation, the use of AI in education should be properly pretested and carefully monitored. Potential issues include data privacy and security, bias and discrimination, and the teacher--student relationship. Certain applications of AI such as personalized learning require students' personal information. For instance, knowing that a student has a learning disability or a mental health issue will allow AI to select the appropriate approach and customize its content accordingly. While students' personal information can be used for tremendous benefit, it can also be susceptible to privacy and security problems. Anonymizing and encrypting the student data will alleviate some of the concerns. However, a comprehensive strategy is required to address this issue. Another important issue is bias and discrimination. Since AI is trained on public data it can be exposed to the biases that exist on the internet. In addition, AI algorithms can also inadvertently learn bias on their own. Since there is a significant amount of entropy in the AI algorithms, their behavior could be unpredictable. Minimizing the amount of bias is one of the key challenges in applying AI in education. 

New technology has historically held potential for misuse. The discovery of nuclear fission created the devastating nuclear bomb. The advent of the internet created the dark web, where illicit and illegal activity can be hidden from the government. However, the society has been able to limit the potential for the abuse of technology through international cooperation and law enforcement. In general, the benefits of new technology outweigh its dangers. Rather than stopping or preventing the advancement of new AI technology in education, it will be more beneficial, on balance, to integrate it into the curriculum. The example of Khan Academy that has recently partnered with OpenAI to integrate ChatGPT into their systems shows a potential roadmap to adopting AI in education. Ultimately, the only way forward is to accept and embrace the new technology, while implementing guardrails to prevent its abuse.

{The main contributions of the paper can be summarized as follows:}
\begin{enumerate}
    \item {Review the existing literature related to AI in education};
    \item {Analyze the potential impact of AI in education};
    \item {Identify the main avenues in applications, benefits, and challenges of AI in education}.
\end{enumerate}

Our paper is organized as follows. Section \ref{sec.2} provides background  details related to ChatGPT. In Section \ref{sec.3}, we present key applications of AI in education. In Section \ref{sec.4}, we discuss the benefits of employing AI in education. Section \ref{sec.5} discusses potential dangers of AI. Section \ref{sec.6} describes future directions and opportunities of AI in education.

\section{ChatGPT \label{sec.2}}
The release of ChatGPT catalyzed the discussion around the benefits as well as dangers of AI. It marked the turning point in the conscience of many people about their perception of AI. While hitherto AI was regarded as mostly a sci-fi fantasy that existed in a far-distant future, the arrival of ChatGPT has suddenly made everyone keenly aware of the legitimacy and tenability of AI. The advent of ChatGPT has increased competition and accelerated the development of alternative AI models motivating the creation of Google's Bard and Meta's LLaMA. In this section, we present a brief background of ChatGPT to provide a more complete picture for the discussion.

\begin{figure}[htp]

\includegraphics[trim={0cm 0cm 0cm 0},clip, width=0.8\textwidth]{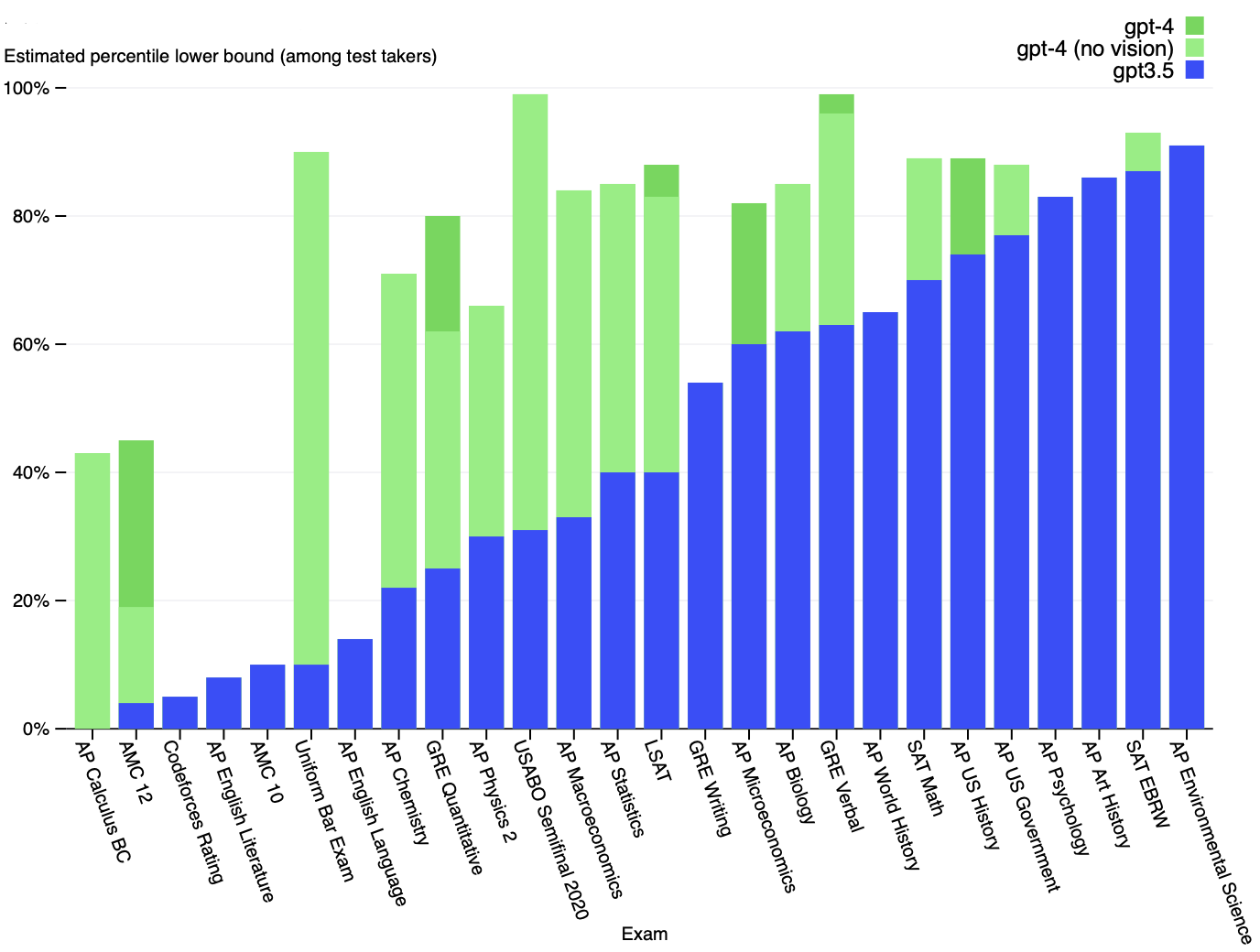}
\caption{Performance of GPT models on various standardized tests.}
\label{chat_gpt}
\end{figure}

ChatGPT is an AI chatbot that was developed by OpenAI. It was initially released on 30 November 2022 based on GPT-3.5 and subsequently updated on 14 March 2023 based on GPT-4 \cite{OpenAI}. It is considered by many as the most powerful AI tool ever crated \cite{Rudolph}. ChatGPT is large language model based on a generative pre-trained transformer (GPT) that is further tuned via supervised and reinforcement learning techniques.  It is able to comprehend and respond to a large variety of prompts with a high level of expertise as well as carry on a continuous dialogue with the user. It can perform a range of tasks from writing poetry in a specified manner and style to generating computer code according to given requirements. While its responses are not perfect, ChatGPT has achieved unprecedented levels of performance. As shown in Figure \ref{chat_gpt}, its latest release based on GPT-4 is able to achieve above-average human performance on several standardized tests including AP tests, SAT, LSAT, and GRE  \cite{GPT}. The success of ChatGPT has been the inflection point in the adoption of AI in society, including education. It has highlighted the abilities of AI and sparked the discourse about its future.

\vspace{-5pt}

\subsection{Transformer}
\begin{figure}[htb]

\includegraphics[trim={0cm 0cm 0cm 0},clip, width=0.5\textwidth]{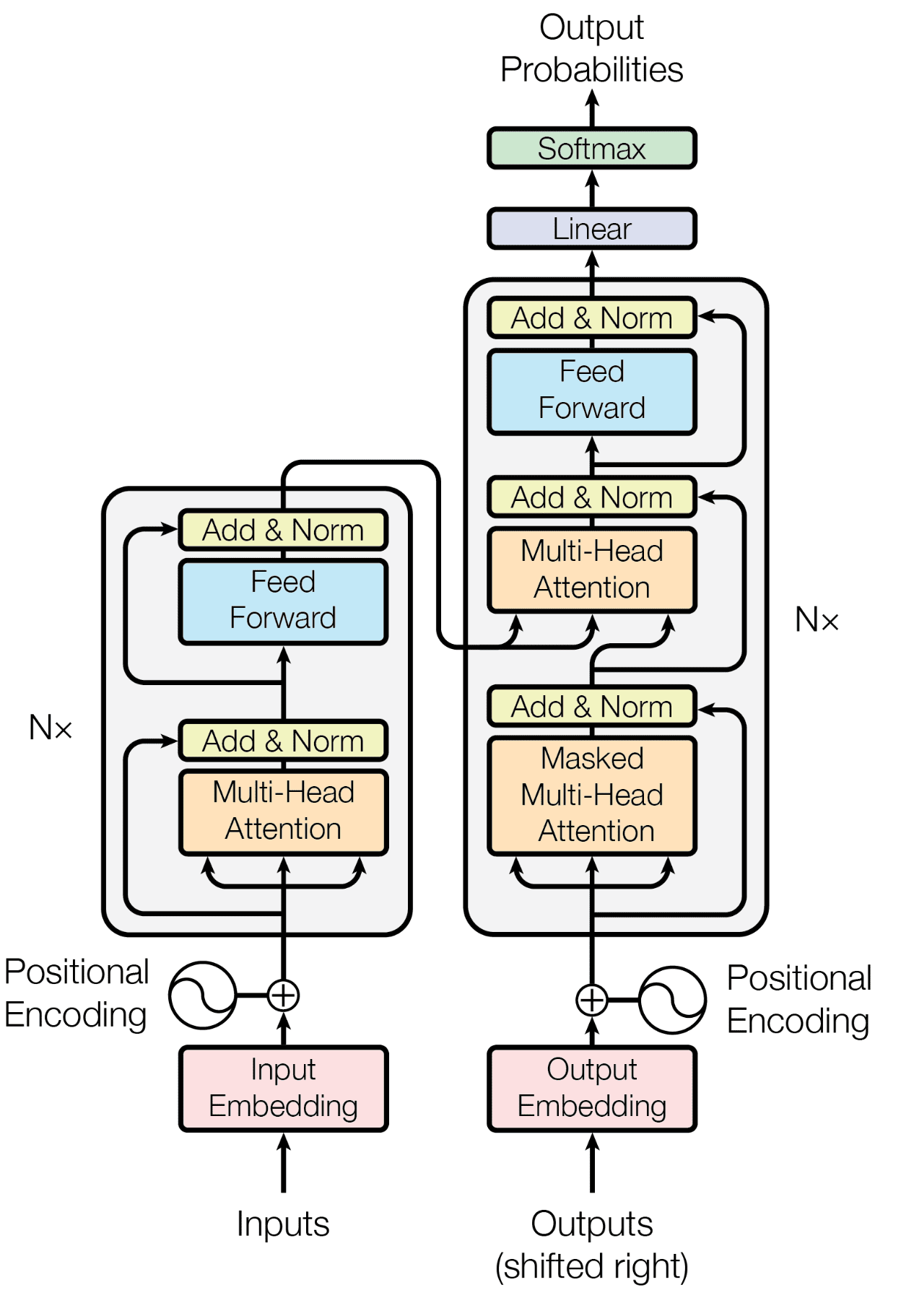}
\caption{The transformer model utilizes the encoder--decoder neural network architecture.}
\label{transformer}
\end{figure}

As with most of the existing large language models including Bard and LLaMA, ChatGPT utilizes the transformer architecture.
Transformer is a sequence-to-sequence neural network model that was originally introduced for language translation but later adopted for general purpose language modeling \cite{Vaswani}. As shown in Figure \ref{transformer}, transformer utilizes an encoder--decoder architecture to build a generative model \cite{Vaswani}.

The key feature of the transformer model is the adoption of self-attention. The attention mechanism provides context for any position in the input sequence, which allows it to process the entire sequence simultaneously. This enables greater parallelization than the recurrent neural networks (RNNs) and therefore reduces training times. Thus, transformers can be trained on large amounts of data which would otherwise be impossible if using RNNs or long short-term memory (LSTM).

The encoder portion of the transformer neural network (Figure \ref{transformer}) is responsible for generating encodings that contain information about which parts of the inputs are relevant to each other. The encoder consist of three major layers: input embedding, multi-head attention, and feed-forward layers. The input embedding layer converts a sequence of input tokens into vectors in the embedding space and adds positional information about each token, which ensures the model can consider the order of elements in the sequence. The embeddings are passed to the attention layer where the model learns various relationships between the input tokens by weighting their importance based on their positions and semantic content. 
The output of the self-attention is passed on to the feed-forward layer, which applies non-linear transformations to further extract higher-level features. These layers are interconnected with residual connections and layer normalization to stabilize and accelerate training. In the end, the encoder outputs rich, context-aware representations for downstream task.

The decoder functions in a reverse manner from the encoder. It takes the output of the encoder and converts it into sequences. The decoder consists of three major layers: masked multi-head self-attention, encoder--decoder attention, and position-wise feed-forward networks.  The masked self-attention mechanism prevents the model from accessing future tokens in the output sequence during training, which ensures that the model is trained autoregressively. The encoder--decoder attention layer then enables the decoder to attend to the encoder's output, allowing it to incorporate the input sequence's contextual information. Finally, the feed-forward networks apply non-linear transformations to refine the generated features using batch normalization and residual connections to improve the training process.

\subsection{GPT}
The transformer architecture gives rise to several popular language models including GPT. In particular, GPT architecture is based on the decoder part of the transformer. 
It is trained to predict the next token in a sequence given the previous tokens in an autoregressive manner, i.e., it is unidirectional, as it processes the input text from left to right, focusing on learning a language model. The distinguishing feature of GPT models is their scale. While no official information has been released regarding GPT-4, the GPT-3 model consists of 175 billion parameters. GPT-3 was trained on 570 GB plaintext and 0.4 trillion tokens using mostly CommonCrawl, WebText, English Wikipedia, and two books corpora (Books1 and Books2). After pretraining, the model was fine-tuned using zero-shot, one-shot, and few-shot learning. It has been speculated that GPT-4 was trained on 1 trillion parameters and cost USD 100 million to train.

\section{Scoping Review \label{sec.3}}
This study aims to review the existing literature with the goal of examining the potential impact of AI in education. As we enter the era of technology-assisted learning, it is important to take stock of the potential applications, benefits, and dangers of AI. We explore the literature to gain insights into various aspects of the problem. 

The review procedure begins with an initial search of the Scopus and Google Scholar databases. We employ appropriate terms and phrases to find the relevant literature. The results are filtered based on various criteria such as quality, applicability, recentness, and others. Then, the screened results are categorized into major themes of applications, benefits, and challenges in AI. The research is further divided into subfields to obtain a more granular view of the subject. 
Finally, we delve into each theme and subtheme discussing in detail the state of affairs. We provide analysis of each subtheme supported by pertinent literature. The main tasks of the study are the following: (1) perform the initial search of Scopus, (2) screen the pertinent works, (3) categorize the information, and (4) provide a detailed discussion of each subtheme.

\subsection*{Methodology}
The research conducted for this article took the form of a scoping review, following the approach described in \cite{Arksey}. Unlike a systematic literature review, a scoping review has broader research {aims} 
 \cite{Schwendimann}. It serves as an effective tool for determining the overall scope and coverage of the existing literature on a particular topic (in the context of this study, an investigation of the effect of AI on education by examining its applications, advantages, and challenges), providing a clear indication of the volume of available literature and studies, as well as an overview of their focus, whether broad or detailed \cite{Munn}. The search process was carried out by two independent researchers between May and June 2023, focusing on the Scopus and Google Scholar databases for the period of 2019--2023. The search yielded a total of 107 results. The study then proceeded through five phases: (1) identifying the research question(s), (2) identifying relevant studies, (3) selecting studies, (4) organizing data, and (5) summarizing and reporting the results \cite{Arksey}).

In \textbf{P{hase 1}
}, the research question and sub-question that were investigated are \mbox{as follows:}
\begin{enumerate}
    \item[RQ1:] How can AI have an impact on Higher Education?
    \item[RQ2:] {What are the benefits and challenges associated with the use of AI in Higher \mbox{Education}}?
\end{enumerate}

\textbf{{Phase 2}} involved the identification of relevant studies. To focus on the most recent research, the database search was limited to the past five years (2019 to June 2023). The search was conducted using Boolean terms, including ``artificial AND intelligence AND positive AND impact AND higher AND education'' (yielding 55 results), ``artificial AND intelligence AND potential AND benefits AND dangers AND higher AND education'' (yielding 0 results), and ``artificial AND intelligence AND applications AND advantages OR disadvantages AND higher AND education'' (yielding 52 results). Abstract and full-text screening were performed by two authors, and the inclusion and exclusion criteria were established by the three authors in agreement. It was argued in \cite{Pollock} that critical appraisal and evaluation of article quality for inclusion in a scoping review were deemed ``non-compulsory''. Assessment of quality, reliability, and confidence using GRADE-CERQual, for instance, was therefore not performed.

\begin{figure}[htb]

  \includegraphics[scale=1]{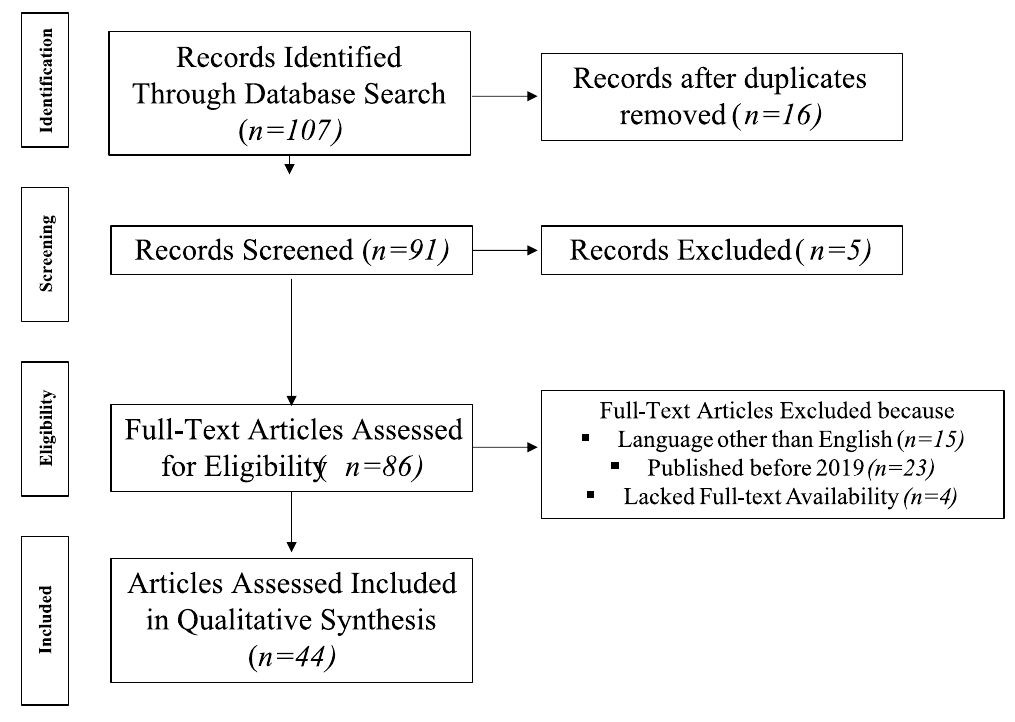}
  \caption{{Overview} 
 of literature search process using PRISMA-ScR.}
  \label{flow_chart}
\end{figure}

During \textbf{{Phase 3}} of the study, a comprehensive selection process was conducted to ensure minimal bias. To maintain consistency, a protocol based on the Preferred Reporting Items for Systematic Reviews and Meta-Analyses guidelines for scoping reviews (PRISMA-ScR) was developed \cite{Tricco}. The selection criteria for the included studies were as follows: (1) written in English, (2) peer-reviewed (articles/book chapters), (3) reports, (4) op-eds, and (5) published between 2019 and June 2023. Studies were excluded if they (a) were published in a language other than English (
15), (b) were published before 2019 (23), and (c) lacked full-text availability (4), as shown in Figure \ref{flow_chart}. A total of 44 articles met the inclusion criteria (refer to Table  \ref{studies}). To ensure consistency and reliability, Krippendorff's alpha coefficient was used to assess inter-rater reliability, resulting in scores of {0}
.81 for abstracts and 1.00 for full texts. Data were shared with the third author. Any disagreements among the three authors regarding study selection were resolved through discussion.

\begin{table}[htb]
\caption{Overview of included {studies}.}
\begin{tabular}{p{0.2cm}p{4cm}p{0.8cm}p{9cm}}
\toprule
&\textbf{Author}&\textbf{Year} & \textbf{Main Contributions of the Surveyed Work} \\
\midrule
  1 & Hleg et al. & 2019 & Presents ethics guidelines for trustworthy AI \\ 
        2 & Terzopoulos et al. & 2019 & Explores the capabilities of voice assistants in the classroom \\

        3 & Webber et al. & 2019 & Discusses the potential application of AI to improve teamwork  \\ 
        4 & Marcinkowski et al. & 2020 & Investigates algorithmic vs. human decision-making in HE admissions \\ 
        5 & Ahmed et al. & 2021 & Investigates an LMS in entrepreneurship \\ 
        6 & Borenstein et al. & 2021 & Examines the ethical implications of Artificial Intelligence technologies \\ 
        7 & González-Calatayud et al. & 2021 & Analyzes the use of AI for student assessment \\ 
        8 & Kamalov et al. & 2021 & Explores Machine Learning for exam-cheating detection \\ 
        9 & Miao et al. & 2021 & Discusses AI in Education policy \\ 
        10 & Timan et al. & 2021 & Discusses data protection in the age of AI \\ 
        11 & UNESCO & 2021 & Provides recommendations on the ethics of AI \\ 
        12 & JISC & 2022 & Discusses and reflects on AI in Tertiary Education \\ 
        13 & Kulshreshtha et al. & 2022 & Explores automatically generated questions as personalized feedback in an ITS \\ 
        14 & Long et al. & 2022 & Explores Collaborative Knowledge Tracing to predict students’ correctness in answering questions  \\ 
        15 & Mishkin et al. & 2022 & Presents risks and limitations for DALL-E 2 \\ 
        16 & Nguyen et al. & 2022 & Discusses ethical principles for AI in Education \\ 
        17 & Oxford Insights Government AI Readiness Index & 2022 & Compares how 160 governments are prepared to use AI in public services \\ 
        18 & Qadir et al. & 2022 & Discusses advantages and drawbacks on Generative AI in education  \\ 
        19 & St-Hilaire et al. & 2022 & Presents the results of a comparative study on learning outcomes for two popular online learning platforms \\ 
        20 & Swiecki et al. & 2022 & Discusses Generative AI and assessment practices  \\ 
        21 & Wahle et al. & 2022 & Explores detection of machine-paraphrased plagiarism  \\ 
        22 & AlAfnan et al. & 2023 & Explores advancements in Artificial Intelligence and its applications \\ 
        23 & Bouschery et al. & 2023 & Focuses on product innovation management research and strategies \\ 
        24 & Chan et al. & 2023 & Presents a preprint discussing specific topics related to teachers, AI and Higher Education \\ 
        25 & Chen et al. & 2023 & Investigates the use of chatbots in classrooms \\ 
        26 & Chetouani et al. & 2023 & Examines human-centered AI, human-centered machine learning, ethics, law, and the societal aspects of AI \\ 

\bottomrule
\end{tabular}
\end{table}

\begin{table}[htb]
 
\caption{{\em cont.}}

\begin{tabular}{p{0.2cm}p{4cm}p{0.8cm}p{9cm}}
\toprule
&\textbf{Author}&\textbf{Year} & \textbf{Main Contributions of the Surveyed Work} \\
\midrule
        27 & Cotton et al. & 2023 & Examines innovations in Education and Teaching  \\ 
        28 & Dai et al. & 2023 & Presents a preprint discussing specific research topics related to education and technology \\ 
        29 & Dwivedi et al. & 2023 & Discusses opportunities as well as ethical and legal challenges of Generative AI \\ 
        30 & Elkins et al. & 2023 & Explores useability of educational questions generated by LLMs \\ 
        31 & European Schoolnet & 2023 & Explores ethical use of digitally processed data for student learning \\ 
        32 & Hu et al. & 2023 & Explores adaptive assessments with Intelligent Tutors \\ 
        33 & Liu et al. & 2023 & Discusses AI and its applications in Education \\ 
        34 & Lodge et al. & 2023 & Discusses the use of Generative AI in tertiary education \\ 
        35 & Liu et al. & 2023 & Presents initial results of a survey on the use if Generative AI at university \\ 
        36 & Malmström et al. & 2023 & Discusses the use of ChatGPT in HE \\ 
        37 & Perkins et al. & 2023 & Discusses academic integrity of LLMs \\ 
        38 & Rasul et al. & 2023 & Presents benefits and challenges of ChatGPT in HE \\ 
        39 & Rudolph et al. & 2023 & Discusses ChatGPT and assessment \\ 
        40 & Sabzalieva et al. & 2023 & Provides an overview of how ChatGPT works and explains how it can be used in HE \\ 
        41 & Sullivan et al. & 2023 & Discusses ChatGPT, academic integrity and student learning \\ 
        42 & UAE & 2023 & Provides a comprehensive guide on the utilization of Generative AI applications \\ 
        43 & Walton Family Foundation & 2023 & Discusses teachers and students’ adoption of Generative AI \\ 
        44 & Wylie et al. & 2023 & Explores the uses of Generative AI in Business Schools  \\ 
\bottomrule
\end{tabular}
\label{studies}
\end{table}


\section{Results \label{sec.4}} 
\textbf{{Phase 4}}: Data from the 44 eligible studies were charted (Table \ref{studies}). Thematic \mbox{analysis \cite{Clarke}} was used to identify themes.

\textbf{{Phase 5}}: Organize and Summarize the Results.
Four overarching themes emerged from the data: (1) Personalized Learning, (2) Intelligent Tutoring Systems (ITS), (3) Assessment Automation, and (4) Teacher--Student Collaboration. To answer RQ 2, benefits and challenges of each of the four themes were identified, as shown in Tables \ref{PersonalizedLearning}--\ref{TeacherStudentCollaboration}.

\begin{table}[htb]

\caption{Overarching themes unpacked: Personalized Learning.}

\begin{tabular}{p{8cm}p{8cm}}
\toprule
\textbf{Benefits} & \textbf{Challenges} \\
\midrule
\begin{itemize}[leftmargin=*]
    \item
      Customized Learning Experience: Personalized learning allows students
      to progress through the curriculum at their own pace, ensuring that
      they fully understand a topic before moving on to the next one. This
      customized approach can lead to better \mbox{learning outcomes}.
    \item
      Improved Student Engagement: When students can learn in a way that
      aligns with their interests, they are more likely to stay engaged and
      motivated, leading to a deeper grasp of the subject matter.
    \item
      Addressing Diverse Learning Needs: Classrooms often consist of
      students with varying levels of knowledge and abilities. Personalized
      learning helps address the diverse learning needs of students,
      providing additional support to those who need it and allowing
      advanced learners to explore more challenging material.
    \item
      Self-directed Learning: Personalized learning encourages students to
      take ownership of their learning and become more self-directed
      learners. They learn how to set goals, monitor their progress, and
      make decisions about their learning path(s).
    \item
      Data-Driven Instruction: Personalized learning relies on data and
      analytics to understand students' strengths and weaknesses. Academics
      can use these data to make informed decisions about instructional
      strategies and interventions.
    \item
      Flexibility and Accessibility: Personalized learning can be
      implemented in various settings and modalities, including classrooms,
      online platforms, and blended learning environments. It also allows
      for greater accessibility for students with special needs, refugee
      contexts, or those in remote areas.
    \item
      Lifelong Learning Skills: By engaging in personalized learning
      experiences, students can develop essential skills and graduate
      outcomes such as critical thinking, problem solving, and independent
      learning, which are valuable beyond their \mbox{academic journey}.
 \end{itemize} &
\begin{itemize}[leftmargin=*]
   \item
      Resource-Intensive: Implementing personalized learning often requires
      significant investments in technology, infrastructure, and continuing
      professional development. Not all institutions may have the resources
      to adopt personalized learning fully.
    \item
      Faculty development and Support: Teachers play a central role in
      personalized learning environments. They need training and ongoing
      support to effectively implement this approach, create personalized
      learning plans, and manage diverse student needs.
    \item
      Curriculum and Content Adaptation: Personalized learning requires a
      curriculum that can be easily adapted to individual learners.
      Developing or curating such flexible content can, however, be
      challenging.
    \item
      Data Privacy and Security: Personalized learning relies on collecting
      and analyzing student data. Ensuring the privacy and security of these
      data is crucial to protect students' information from potential
      breaches or misuse.
    \item
      Technical Issues: The use of technology in personalized learning can
      be susceptible to technical glitches and interruptions. Reliability
      and seamless integration of technology are essential for a smooth
      personalized learning experience.
    \item
      Balancing Structure and Freedom: Some students may struggle with the
      freedom to choose their learning path and may need academic advising.
    \item
      Assessment and Accountability: Traditional forms of standardized
      assessment may not align well with personalized learning approaches.
      Finding effective ways to assess (summative and formative) and measure
      student progress in personalized settings can be rather complex.
    \item
      Equity Concerns: Ensuring equitable access to personalized learning
      opportunities is essential. Some students may face barriers to access,
      such as limited internet connectivity or access to devices, which
      could exacerbate existing achievement gaps.
\end{itemize}\\
\bottomrule
\end{tabular}
\label{PersonalizedLearning}
\end{table}

\begin{table}[htb]

\caption{Overarching themes unpacked: Intelligent Tutoring Systems.}
\begin{tabular}{p{8cm}p{8cm}}
\toprule
\textbf{Benefits} & \textbf{Challenges} \\
\midrule
\begin{itemize}[leftmargin=*]
   \item
      Enhanced Learning Outcomes: ITS can provide personalized and adaptive instruction, tailored to individual student needs, which can lead to improved learning outcomes and academic performance.
    \item
      Individualized Learning: ITS can offer personalized feedback and
      guidance to students, allowing them to progress at their own pace and
      focus on areas where they need additional support.
    \item
      Continuous Assessment: ITS can provide real-time assessment and
      feedback, enabling students to monitor their progress and identify
      areas for improvement.
    \item
      Immediate Feedback: ITS can offer immediate feedback on students'
      responses, allowing them to correct their mistakes and reinforce their
      understanding in real time.
    \item
      Access to Quality Education: ITS can provide access to quality
      education in remote or underserved areas, reaching students who may
      not have access to traditional educational resources.
    \item
      Cost and Time Efficiency: ITS can reduce the cost and time associated
      with one-on-one tutoring by automating certain aspects of instruction
      and support.
    \item
      Multimodal Learning: ITS can incorporate various forms of multimedia,
      such as videos, interactive simulations, and virtual environments, to
      engage students and enhance their learning experience.
    \item
      Long-term Knowledge Retention: ITS can employ spaced repetition and
      other cognitive techniques (Cognitive Load Theory) to promote
      long-term retention of knowledge and skills.
\end{itemize} &
\begin{itemize}[leftmargin=*]
    \item
      Development and Implementation: Designing and developing effective ITS requires significant time, training, resources, and expertise in both pedagogy and technology.
    \item
      Data Privacy and Security: ITS collect and analyze large amounts of
      student data, raising concerns about data privacy, security, and the ethical use of personal information.
    \item
      Bias and Discrimination: If not properly designed and trained, ITS can perpetuate biases and discrimination, as the underlying algorithms may replicate biases present in the training data.
    \item
      Technical Limitations: ITS may face limitations in accurately
      understanding and interpreting students' responses, especially in
      complex or ambiguous situations, leading to potential gaps in
      instructional support.
    \item
      Teacher--Student Relationship: The use of ITS may impact the
      traditional teacher--student relationship, as students may rely more on the system for instruction and guidance, potentially reducing interpersonal interactions.
    \item
      User Acceptance and Engagement: ITS may face resistance or low user
      acceptance from students and academics, who may prefer traditional
      instructional methods or perceive the system as impersonal or less
      effective.
    \item
      Lack of Adaptability: Some ITS may struggle to adapt to individual
      learning styles, preferences, and cultural differences, potentially
      limiting their effectiveness across diverse student populations.
    \item
      Integration with Existing Systems: Integrating ITS with existing
      educational technologies, infrastructure, and curricula can present
      technical and logistical challenges, requiring careful planning \mbox{and
      coordination.}

\end{itemize}\\
\bottomrule
\end{tabular}
\label{ITS}
\end{table}

\begin{table}[htb]
\caption{Overarching themes unpacked: Assessment Automation.}
\begin{tabular}{p{8cm}p{8cm}}
\toprule
\textbf{Benefits} & \textbf{Challenges} \\
\midrule
\begin{itemize}[leftmargin=*]
\item
  Time Efficiency: Automation can significantly reduce the time and
  effort required for grading and evaluating assessments. Computer-based
  grading systems can quickly score multiple-choice questions,
  fill-in-the-blank responses, and even some types of open-ended
  questions, saving academics valuable time.
\item
  Consistency and Reliability: Automated grading ensures consistent and
  objective evaluation, eliminating potential bias or human error. Each
  student's assessment is evaluated against the same criteria, promoting
  fairness and accuracy in the \mbox{grading process}.
\item
  Faster Feedback: Automation allows for faster feedback delivery to
  students. Instead of waiting for manual grading, students can receive
  instant feedback on their assessments, enabling them to identify areas
  of improvement and adjust their learning \mbox{strategies promptly}.
\item
  Scalability: Automation enables the grading of a large number of
  assessments efficiently. It is particularly beneficial for online
  courses, MOOCs, or programs with a high volume of assessments, as it
  can handle large student populations without sacrificing the quality
  of evaluation.
\item
  Analytics and Insights: Automated assessment systems can generate data
  and analytics on student performance, allowing academics and
  institutions to gain insights into student learning patterns, identify
  common misconceptions, and make data-driven instructional decisions.

\end{itemize} &
\begin{itemize}[leftmargin=*]
\item
  Limited Applicability: Not all types of assessments can be easily
  automated. While diagnostic assessment, multiple-choice questions, and
  some structured responses lend themselves well to automation,
  subjective or complex assessments requiring human judgment, such as
  essays or project evaluations, may be challenging to automate fully.
\item
  Adaptability: Automated assessment systems may struggle with adapting
  to unique or creative student responses that deviate from predefined
  answer patterns/rubrics. They may not be able to recognize innovative
  or unconventional thinking, limiting the scope of assessment.
\item
  Learning Outcomes Assessment: Some learning outcomes, such as critical
  thinking, creativity, and problem solving, are difficult to assess
  through automated systems alone. These higher-order skills often
  require human judgment and qualitative evaluation, which automated
  assessments may not capture adequately.
\item
  Technical Limitations: Assessment automation relies on technology, and
  technical issues can arise, such as system errors, software glitches,
  or compatibility problems. These technical limitations can disrupt the
  assessment process and impact its reliability and validity.
\item
  Lack of Contextual Understanding: Automated systems may struggle to
  understand the context or nuances of student responses, leading to
  potential misinterpretation or incomplete evaluation. They may not
  grasp the underlying reasoning or provide targeted feedback to address
  individual student needs effectively.
 \end{itemize}\\
\bottomrule
\end{tabular}
\label{AssessmentAutomation}
\end{table}
  
\begin{table}[htb]
\caption{\em{cont}.}
\begin{tabular}{p{8cm}p{8cm}}
\toprule
\textbf{Benefits} & \textbf{Challenges} \\
\midrule
\begin{itemize}[leftmargin=*]
\item
  Personalization: Automated assessment tools can be designed to provide
  personalized feedback and recommendations based on individual student
  performance. This customization helps students understand their
  strengths and weaknesses, guiding them toward targeted learning
  activities.
\item
  Standardization: Automated assessments can be aligned with
  predetermined program/course learning outcomes, ensuring that students
  are evaluated consistently against specific educational standards.
  This standardization helps maintain the quality and integrity of
  assessments and helps ensure constructive alignment.
\end{itemize} &
\begin{itemize}[leftmargin=*]
\item
  Teacher--Student and Student--Student Interaction(s): Automated
  assessments, particularly those lacking human involvement, can limit
  opportunities for meaningful interaction(s) and dialogue between
  teachers and students and students with other students (peer
  assessment). Personalized feedback and guidance may be absent,
  impacting the overall learning experience.
\item
  Ethical Considerations: Assessments that involve sensitive or personal
  information require careful handling and protection of student data.
  The automation of assessments raises ethical concerns related to data
  privacy, security, and potential biases in algorithms or automated
  decision-making processes.
\end{itemize}\\
\bottomrule
\end{tabular}
\label{AssessmentAutomation2}
\end{table}

\begin{table}[htb]
\caption{Overarching themes unpacked: Teacher--Student Collaboration.}
\begin{tabular}{p{8cm}p{8cm}}
\toprule
\textbf{Benefits} & \textbf{Challenges} \\
\midrule
\begin{itemize}[leftmargin=*]
\item
  Personalized Learning: AI tools can analyze student data and provide
  personalized recommendations and resources based on individual
  learning needs. This personalization helps academics tailor their
  instruction to each student, fostering a more effective and targeted
  learning experience.
\item
  Real-time Feedback: AI tools can provide instant feedback to students
  on their work, allowing them to identify mistakes, misconceptions, or
  areas needing improvement promptly. This immediate constructive
  feedback helps students make adjustments in their learning strategies
  and promotes continuous learning and growth.
\item
  Enhanced Communication: AI tools facilitate communication and
  collaboration between teachers and students through various channels.
  These tools can support online discussions, virtual classrooms, and
  interactive platforms, enabling seamless interaction and engagement
  regardless of physical location, for both the academic and the
  student.
\item
  Resource Accessibility: AI-powered platforms can offer a wide range of
  educational resources, including e-books, multimedia materials, and
  interactive simulations. These tools provide students with access to
  diverse learning materials and enable teachers to share resources
  easily, expanding the learning opportunities beyond traditional
  classroom boundaries.
\item
  Data Analysis and Insights: AI tools can analyze large volumes of
  student data, such as performance, behavior, and
  navigational/engagement patterns. This data analysis provides valuable
  insights to academics, enabling them to track student progress,
  identify trends, and make data-driven instructional decisions for
  individual students and/or the entire class.
  
\end{itemize} &
\begin{itemize}[leftmargin=*]
\item
  Lack of Human Interaction: AI tools, while beneficial, cannot fully
  replace human interaction and the value of face-to-face communication
  between teachers and students. Over-reliance on AI tools may result in
  reduced opportunities for meaningful interactions, personal
  connections, sense of belonging, and emotional support.
\item
  Technical Issues and Reliability: AI tools rely on technology, and
  technical glitches or system failures can disrupt the collaboration
  process. Reliability and stability issues may affect the trust and
  confidence in AI tools, potentially impacting their adoption and
  effectiveness.
\item
  Data Privacy and Security: AI tools often require the collection and
  analysis of student data, raising concerns about privacy and security.
  Proper protocols and safeguards must be in place to protect sensitive
  student information and ensure compliance with data protection
  regulations.
\item
  Bias and Fairness: AI tools are only as unbiased as the algorithms and
  data they are trained on. If the underlying data or algorithms have
  biases, it can lead to unfair outcomes or perpetuate existing
  inequalities. It is crucial to regularly evaluate and mitigate biases
  to ensure fairness and equity in the use of AI tools.
\item
  Skills and Training: Effective utilization of AI tools requires
  teachers and students to have the necessary skills and training.
  Academics need support and professional development opportunities to
  understand and leverage AI tools effectively. Similarly, students need
  guidance on how to use AI tools appropriately and critically evaluate
  the information provided.
\end{itemize}\\
\bottomrule
\end{tabular}
\label{TeacherStudentCollaboration}
\end{table}

\begin{table}[htb]
\caption{\em{cont.}}
\begin{tabular}{p{8cm}p{8cm}}
\toprule
\textbf{Benefits} & \textbf{Challenges} \\
\midrule
\begin{itemize}[leftmargin=*]
\item
  Time Efficiency: AI tools can automate routine administrative tasks,
  such as grading assignments or organizing schedules, allowing
  academics to dedicate more time to instruction and personalized
  support. This time efficiency frees up academics' workload and enables
  them to focus on higher-value teaching and learning activities.
\end{itemize} &
\begin{itemize}[leftmargin=*]
\item
  Ethical Considerations: The use of AI tools raises ethical
  considerations, such as the responsible use of student data,
  transparency in algorithms and decision-making processes, and ensuring
  that AI tools do not compromise student well-being or autonomy.
\end{itemize}\\
\bottomrule
\end{tabular}
\label{TeacherStudentCollaboration2}
\end{table}

The papers suggest that AI has the potential to positively impact Higher Education. As was argued in \cite{Popkhadze}, for instance, that AI, big data, and learning analytics can become a powerful tool for advancing higher education institutions further, at the same time, AI can have a detrimental effect without a vigilant eye. However, the papers also acknowledge the challenges and potential risks of introducing AI in educational settings, such as the need to pay attention to the ethical side of AI.

\begin{figure}[htb]

\includegraphics[trim={0cm 0cm 0cm 0cm},clip, width=1\textwidth]{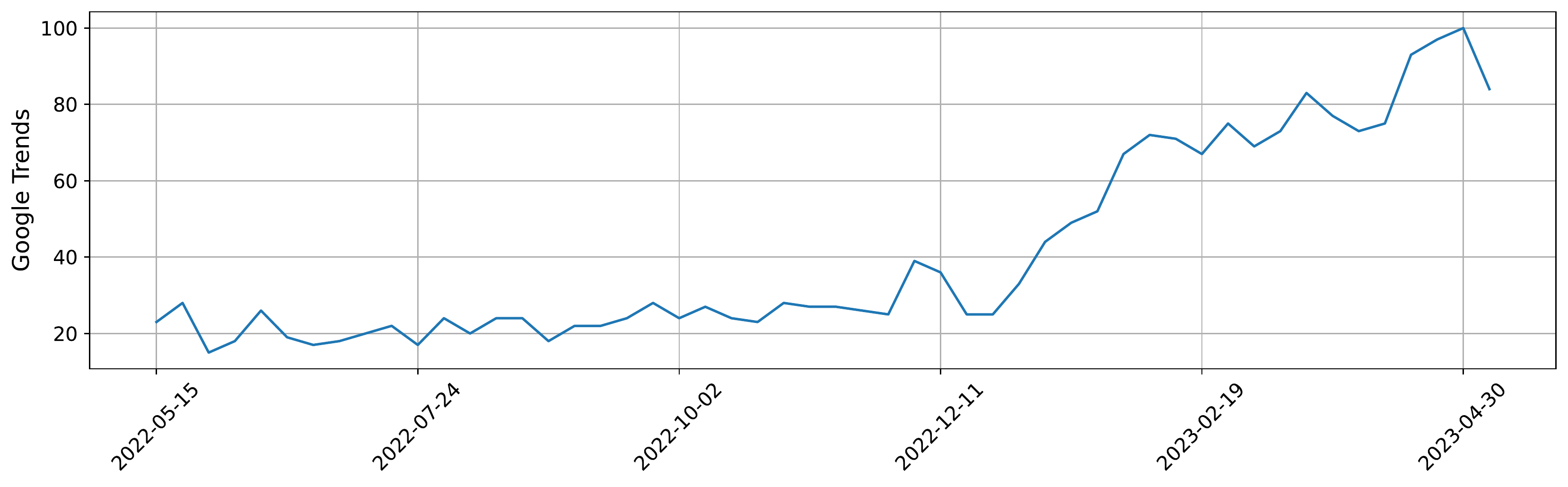}
\caption{Popularity worldwide of the search query ``AI in education'' \cite{trends}. }
\label{popularity}
\end{figure}

Our review reveals that the recent popularity of AI in education (Figure \ref{popularity}) is driven by four main applications: (1) Personalized Learning, (2) Intelligent Tutoring Systems, (3) Assessment Automation, and (4) Teacher--Student Collaboration.

\subsection{Theme 1: Personalized Learning}
Personalized learning powered by AI involves tailoring educational content and experiences to cater to each student's unique needs, learning styles, and preferences. This approach enables students to learn at their own pace, thereby enhancing their engagement and overall learning outcomes \cite{Luan}. 
The individualized approach benefits both students and teachers, leading to better learning results and overall educational experiences. AI has the ability to help tailor education in a variety of ways. Adaptive learning systems driven by AI may analyze students' performance, strengths, and shortcomings in order to offer tailored learning courses. AI algorithms and adaptive learning systems can analyze student data, recognize patterns, and suggest personalized content and resources to optimize their learning experience. Based on each student's achievement, these platforms may change the speed, material, and complexity of the curriculum, to deliver an optimal learning experience. AI can power virtual tutors who provide students with one-on-one education customized to their unique learning and emotional needs \cite{Jonnalagadda, Mahfood}. Automated tutoring systems can provide instant feedback, answer queries, and walk pupils through complex ideas, supplementing or even replacing traditional tutoring services.

AI has the potential to analyze massive volumes of data created by students during the learning process, revealing patterns and trends that might assist instructors in identifying areas where students require additional support or resources. It enables instructors to develop individualized learning experiences for the students based on data.
The power of AI can help improve learning experiences by including gamification features such as incentives, challenges, and competition. Engaging and individualized learning environments can be developed that increase motivation and active engagement by adapting these components to individual students.
AI has potential to help students with special needs or impairments receive tailored education by designing assistive technologies that meet their specific demands. Speech recognition and text-to-speech technologies can be employed to develop custom content that fits the individual needs of affected students.

\subsection{Theme 2: Intelligent Tutoring Systems}

\begin{figure}[htb]

\includegraphics[trim={0cm 0cm 0cm 0cm},clip, width=0.6\textwidth]{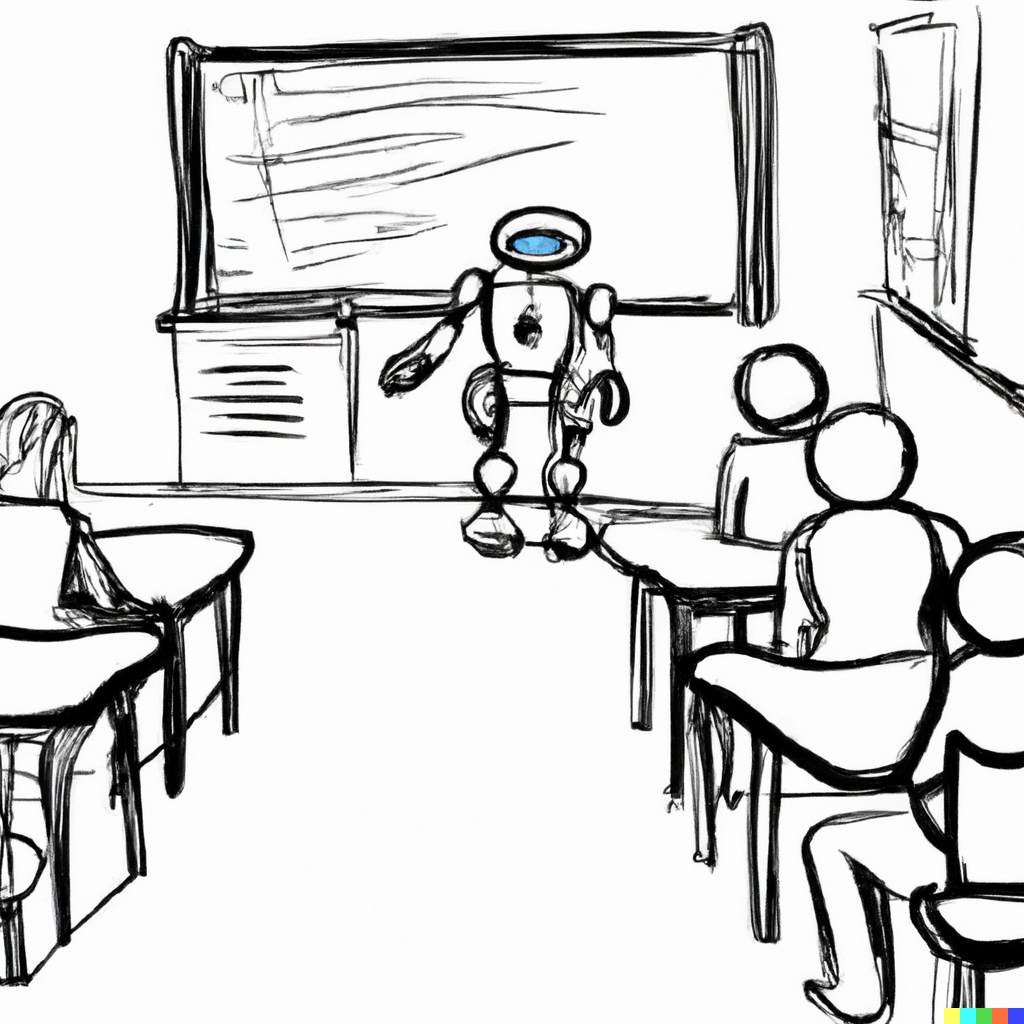}
\caption{Intelligent humanoid leading a tutoring session (DALL·E 2 image). }
\label{its}
\end{figure}

AI technology has the potential to accelerate intelligent tutoring systems (ITS), which are aimed at providing students with tailored, one-on-one teaching, emulating the experience of learning from a human tutor (Figure \ref{its}). To understand students' learning needs and customize their teaching methods, ITS employs powerful algorithms and machine learning approaches. Natural language processing (NLP) enables AI to perceive and interpret written or spoken input from students, allowing ITS to engage in meaningful dialogues, answer questions, and provide instruction on a variety of subjects.
ITS employs AI to provide customized instruction and feedback to students, thus bridging the gap between traditional classroom learning and individualized tutoring \cite{Mousavinasab}. 

Student modeling is another critical component of ITS. It is a process in which AI generates dynamic models of students' knowledge, skills, and learning preferences based on their interactions with the system. These models assist ITS in understanding students' present comprehension of the material, misconceptions, and areas where they may require additional help. 
These systems analyze students' learning behaviors, track their progress, and offer personalized guidance, contributing to improved learning outcomes and increased student motivation \cite{Kochmar}. AI may develop ITS that customizes education to individual learners and successfully addresses their specific needs by combining NLP, student modeling, and adaptive content delivery.

\subsection{Theme 3: Assessment Automation}
Automated student assessment is one of the major areas of application.
AI has the potential to significantly improve student assessment by automating it, expediting the evaluation process, and providing fast, individualized feedback \cite{Minn}. It can revolutionize the way assessments are conducted by providing real-time feedback to students \cite{Gonzalez}.  AI can examine students' work, from simple multiple-choice questions to sophisticated written assignments and even spoken responses, by employing machine learning techniques and NLP. AI-driven assessment tools can evaluate complex tasks, such as essays and problem-solving exercises, reducing the burden on educators while improving the accuracy and consistency of grading \cite{Vittorini}. This feature enables more objective evaluation, eliminating human bias and errors while allowing educators to focus on other areas of teaching, such as curriculum creation and student engagement. The technology has fewer time and resource constraints compared to a human educator, which allows AI to provide more in depth feedback on student work. Since learning from mistakes is one of the most effective ways of learning, high-quality feedback is important in mastering the material.

Automated AI assessment tools can detect patterns in students' performance, identifying specific areas where they may require additional assistance or resources. It helps instructors to make data-driven decisions and build personalized interventions that suit individual learning needs by offering real-time feedback and deep insights into students' strengths and weaknesses. This level of personalization and adaptation in evaluation can help students achieve better learning outcomes and have a more engaging \mbox{educational experience}.

\subsection{Theme 4: Teacher--Student Collaboration}

The collaboration between teachers and students can be facilitated by AI, enhancing the overall learning experience. By providing real-time analytics and insights, AI can help educators identify students' strengths, weaknesses, and learning patterns, allowing them to adjust their teaching strategies accordingly \cite{Villegas}. In situ assessments and instant feedback allow teachers to make real-time adjustments during the class. AI can be used to notify teachers when students, and which students, are struggling while providing possible remedies. As a brainstorming partner, AI can help identify effective solutions to support student learning.

AI technology can help teachers to answer a variety of student questions in class. Since human teachers have a limited amount of knowledge, they may be challenged by unexpected and out-of-box questions from the students. AI can help fill the gap in knowledge and provide high-quality responses to student questions.
AI-powered chatbots can address students' queries and facilitate peer-to-peer interactions, promoting a collaborative learning environment \cite{Lee}. The AI-assisted approach has been particularly popular in English as a second language (ESL) courses \cite{Klimova, Lin}.

Voice-activated technology that uses speech recognition technology is another area of advancement that will help improve teacher--student collaboration.

\section{Advantages of AI in Education \label{sec.5}}

AI technology holds tremendous potential for enhancing the quality of education at all levels of study. The key advantages of AI can be summarized as follows: (1) enhanced learning outcomes, (2) time and cost efficiency, and (3) global access to quality education.

\subsection{Enhanced Learning Outcomes}

AI in education has the capacity to significantly improve student learning outcomes by providing tailored learning experiences with the help of knowledge tracing (KT) and collaborative filtering. AI technology can be used to customize learning experiences to match and predict specific student needs (and future performance) by (a) identifying individual strengths and limitations, (b) tracing knowledge states \cite{Long}, and (c) analyzing past interactions and navigational patterns. As a result, more engaged and motivated students, increased information retention, and, ultimately, enhanced academic success may be achieved. Patterns and trends in student data (actionable intelligence) can be identified with the assistance of AI to help educators identify and address any learning gaps or challenges in real time. A more refined and granular adaptive exercise recommendation based on (students’) cognitive level and collaborative filtering with an 8 percent improvement in recommendation effectiveness was proposed in \cite{Liu23}. The potential of intelligent tutoring systems to enhance learning outcomes by delivering tailored instruction and just-in-time feedback to \mbox{students \cite{Kulshreshtha}} was recognized as early as 1990 \cite{Nwana}. More recently, several studies found that by catering to individual learning styles and preferences, AI can increase student engagement, motivation, and retention \cite{Bhutoria, Hao} with the use, for instance, of personalized feedback systems based on deep learning-based Transformer models \cite{Kulshreshtha}  or Intelligent Tutoring Systems/AI-powered Tutoring (e.g., Chrome’s Galileo or Khanmigo). A recent study \cite{StHilaire} indicated, for instance, that learning outcomes were “better for participants on Korbit”, a Canadian AI-powered learning platform, which uses machine learning models to adapt the learning process to students, “than participants on either of the platforms that” did “not provide personalized feedback”. Learning outcomes can also be improved by automating often time-consuming administrative tasks like assessment, grading, and teaching and learning activity planning, allowing educators to focus on direct student interaction and effective teaching strategies. Existing technology-driven instruction \mbox{tools \cite{Kamalov22b, Kamalov23}} can also be further enhanced with the power of AI.

AI tools and resources, such as virtual tutors, voice assistants \cite{Terzopoulos}, text-to-image generation (Stable Diffusion, Dall-e-2, Midjourney),  text summarization (summarizer.org),  AI video generation platforms (Synthesia or Elai.io), AI-enabled adaptive assessment \cite{Hu}, and smart content, may also supplement traditional learning techniques by providing students with additional material and interactive experiences that improve their understanding of complex concepts \cite{Chan, Qadir}. The Joint Information Systems Committee (JISC) interactive map shows how UK tertiary institutions use AI technologies to improve processes. The ChatGPT quick start guide by Sabzalieva and Valentini \cite{Sabzalieva} and the United Arab Emirates 100 practical applications and uses of generative AI \cite{UAE} provide useful and valuable examples of how AI can be used to improve productivity and facilitate student collaboration by promoting teamwork, team effectiveness \cite{Webber}, and problem-solving skills, all of which are required to succeed in today’s workforce. Teamwork during the class is a commonly used teaching strategy that helps students to work together and learn from one another. Since teachers can attend to only one team at a time, AI can be employed to assist and lead the discussion in each group, in a “hybrid team”, according to intent and topics (intelligent triage). AI can help steer teamwork in the right direction. The researchers in \cite{Bouschery}  explored how GPT-3 could act as an innovator in a hybrid (augmented) team in the new product development process. In general, incorporating artificial intelligence into education has the potential to create a more engaging, efficient, and effective learning environment that benefits both students and educators. 

\subsection{Time and Cost Efficiency}

\begin{figure}[H]

\includegraphics[trim={0cm 0cm 0cm 0cm},clip, width=0.9\textwidth]{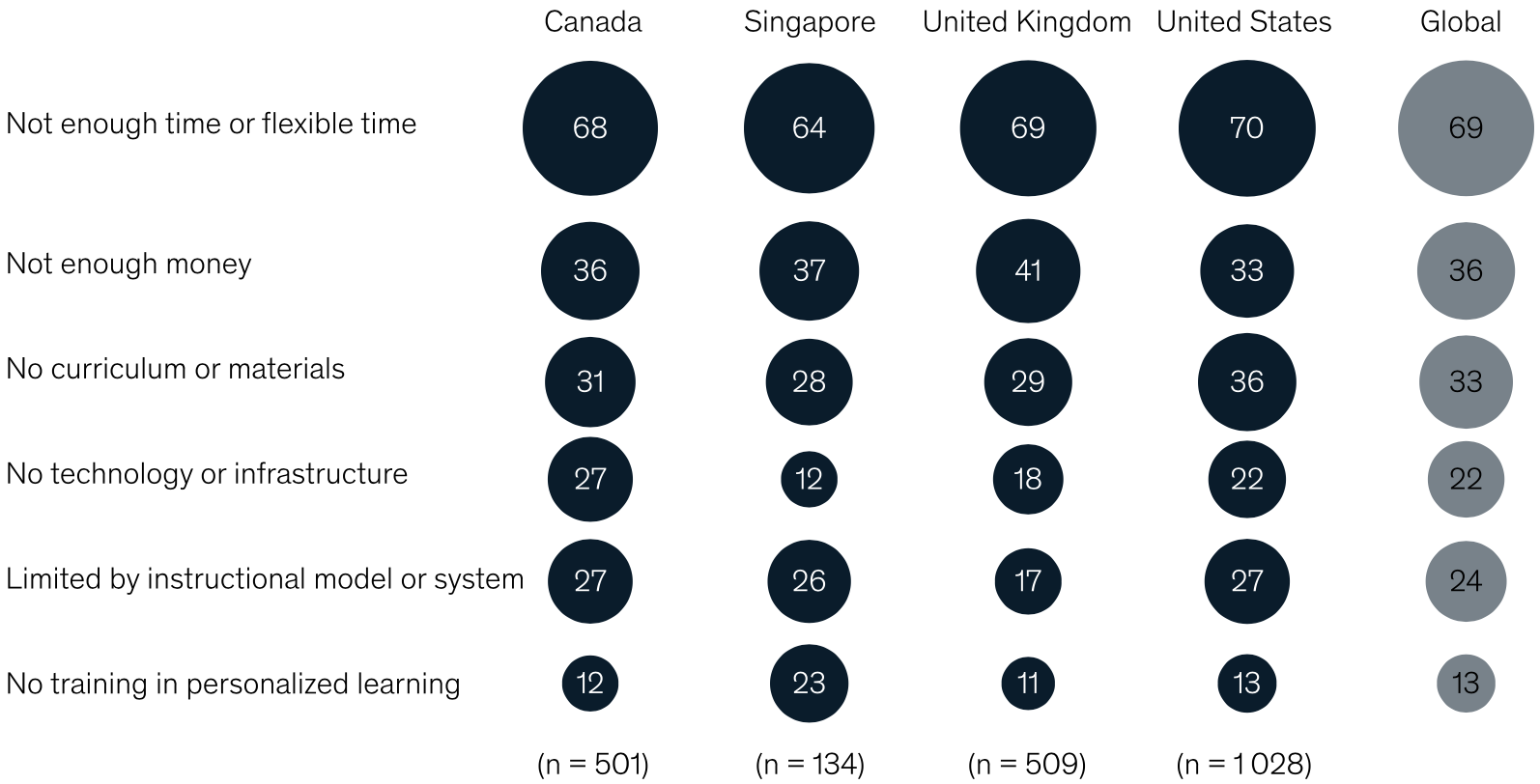}
\caption{{Top} 
 barriers to providing personalized learning, \% of teachers identifying area as a \mbox{primary barrier}.}
\label{personalizedlearning}
\end{figure}

One of the key advantages of AI is the automation of manual tasks \cite{Swiecki}. The introduction of AI in education has resulted in a paradigm shift, with potential to dramatically improve time and cost efficiency for both students and educators. The automation of teachers' duties will provide time savings that would allow greater focus of personalized learning (Figure \ref{personalizedlearning}) \cite{mckinsey}.

AI can streamline various educational processes such as generating questions using prompts \cite{Elkins}, grading, and content creation, reducing the workload on educators and enabling them to focus on higher-order tasks including thinking and problem solving. Specifically, AI can be trained to grade a range of subjects and assessments including essays, problem-solving questions, and even graphing assessments \cite{Gonzalez}. Since grading consumes up to half of educators' time, the potential time savings are enormous. In addition, the instant grading provided on assessments ensures that students receive timely and constructive feedback on their performance \cite{Dai, Lodge}. AI-Generated Lesson Plan creation is another avenue for improved efficiency \cite{Ahmed}. AI can be used to create lecture slides, custom images and videos, homework and exam questions, and other course content, allowing academics to focus on human-centric tasks. AI tools can provide cost-effective solutions for institutions, particularly in resource-constrained settings, by offering scalable and accessible learning opportunities \cite{Luan2020}. 

AI has the potential to alter educational institution budget allocations by reducing reliance on traditional, resource-intensive teaching methods in terms of cost efficiency. With the incorporation of AI into the educational scene, schools can save money on hiring many teachers and other staff while still providing a high-quality education to their pupils. AI-based educational resources such as Massive Open Online Courses (MOOCs) and digital textbooks can be delivered at a fraction of the cost of their traditional counterparts. As a result, these low-cost frugal solutions and innovations improve access to education, especially for students in low-resource settings or in remote {regions} 
 \cite{Santandreu2019}, democratizing information and contributing to the wider educational setting. In the new paradigm, the role of the educator will be to oversee the work of AI. Quality control and fine-tuning the AI will become key responsibilities of a teacher. While AI can be utilized to perform various tasks, educators must approve the results.

\subsection{Global Access to Quality Education}
\begin{figure}[H]

\includegraphics[trim={0cm 0cm 0cm 0cm},clip, width=0.8\textwidth]{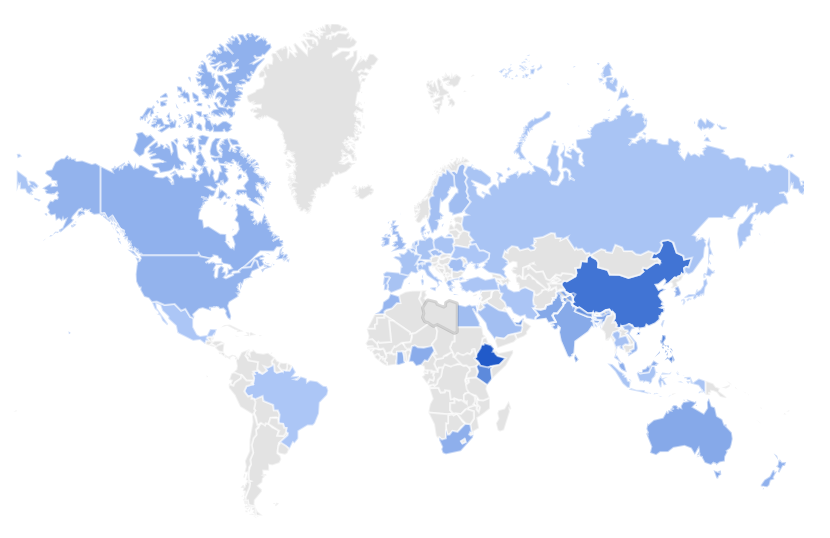}
\caption{Popularity of the search query ``AI in education'' by region. }
\label{regions}
\end{figure}

The lack of access to quality education is one of the key issues in many developing nations \cite{Shah2023}. It is a sentiment that is expressed by the leading experts in international education \cite{EuropeanSchoolnet}. Since education is an important factor in driving economic growth, providing universal education has huge financial benefits to both individuals and countries. The emergence of AI has paved the way for significant breakthroughs in worldwide access to high-quality education. As a result, AI-based education has recently attracted a significant amount of interest from around the globe (Figure \ref{regions}) \cite{trends}. AI-powered educational tools and materials can cross geographical, socioeconomic, and linguistic boundaries, allowing for a fairer distribution of information. 
{In this regard, it is important to consider the technological challenges of delivering AI-based education to remote locations. Thus, the next-generation communication networks, such as 6G, play a crucial role in enabling the above vision} \cite{Khanh}.

One of the key advantages of AI is its relatively low cost. While the initial fixed cost of building an AI system can be significant, as in the case of GPT-4, which was estimated to cost USD 100 million to train, the variable cost is almost zero. In other words, AI systems can scale at low cost. AI has the potential to democratize education by providing access to high-quality learning resources and personalized instruction across geographic and socioeconomic barriers \cite{Tondeur}. Online AI-driven platforms and chatbots can bridge the gap between students and quality education, helping to reduce inequalities and create a more inclusive learning environment  \cite{Huang}. Since English presentation skills are still a major barrier for some participants, ChatGPT can help tell a more convincing story when pitching their ideas in front of the jury, as well as written reports. It means an equal playing field for participants from all backgrounds \cite{Wylie}. By adapting educational content and pedagogical approaches to meet the needs and abilities of different communities, AI-powered adaptive learning systems can improve students’ engagement, motivation, and academic achievement, making education more effective and accessible around \mbox{the world}.

\section{Challenges and Ethical Considerations \label{sec.6}}
\begin{figure}[H]

\includegraphics[trim={0cm 0cm 0cm 0},clip, width=1\textwidth]{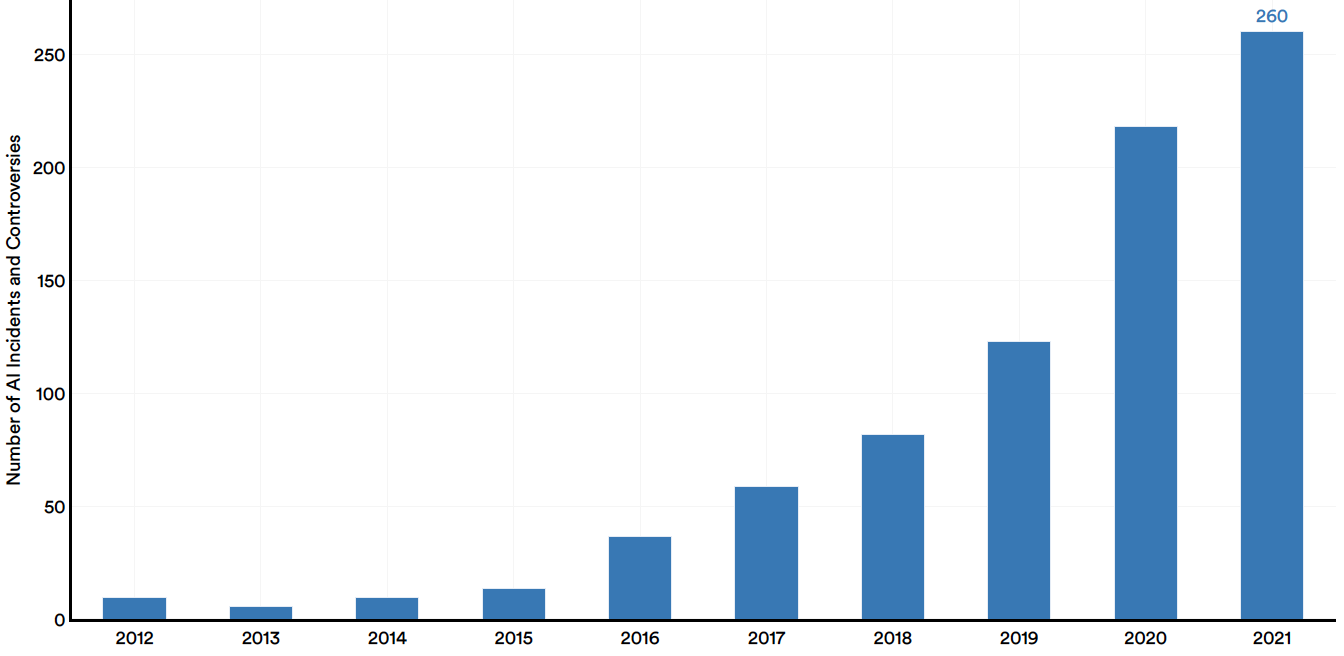}
\caption{Number of ethics incidents related to AI.}
\label{ethics}
\end{figure}

While AI has potential to revolutionize the education system, there are several concerns over its dangers \cite{Holmes}. As shown in Figure \ref{ethics}, the number of ethics incidents has continued to grow rapidly in line with the rise of AI  \cite{Maslej}.
Several AI ethics initiatives have emerged recently \cite{AI, Universite}. However, very few schemes currently exist for the specific issues raised by AI in education \cite{Holmes}, except maybe the European Commission Ethics guidelines for trustworthy AI \cite{Hleg} and the forthcoming UNESCO IESALC Manual on AI in Higher Education. It is important to move carefully in the adoption of AI in schools and universities. There is a real danger of AI becoming pervasive in every sense where those involved may be exposed to risks without being aware of them \cite{Nguyen, Nazir}. Although governments have recognized the importance of AI for future development, there is lack of comprehensive policy or guidance around AI in education \cite{Chetouani, Miao, Schiff}.

Proper testing and evaluations are needed to ensure the safety of the technology. The use of AI with young children should be particularly investigated. The potential issues related to AI include the following: (1) data privacy and security, (2) bias and discrimination, (3) plagiarism and academic integrity, and (4) the teacher--student relationship.

\subsection{Data Privacy and Security}
Data privacy and security concerns affect almost every technology. User personal data, communications, and location details are vulnerable to hacking or monitoring by governments. These concerns also exist in the applications of AI within the educational domain. As AI tools penetrate educational institutions, they collect and process vast quantities of sensitive information, including students’ personal data, academic records, and behavioral patterns. Consequently, the potential for misuse or unauthorized access to these data presents significant risks to stakeholders \cite{Timan}. The violation of individual privacy rights and the potential for discriminatory practices are among key dangers. In addition, inadequate data security measures may lead to breaches or leaks, further exacerbating these concerns and undermining trust in AI-based educational systems. The implementation of AI in education raises concerns regarding data privacy and security, as these systems often rely on large volumes of student data for analysis and personalization \cite{Miao, Zeide}. The challenge lies in striking a delicate balance between harnessing AI’s potential to revolutionize education and ensuring the protection of data privacy and security. Various solutions have been proposed, including the implementation of robust data protection policies, encryption techniques, and stringent access controls. Ensuring the protection of sensitive information and complying with data privacy regulations, such as the General Data Protection Regulation (GDPR), is crucial to maintain trust and prevent potential misuse \cite{Bessen2020}. Moreover, fostering a culture of privacy awareness within educational institutions is critical to ensure compliance with data protection laws and regulations. It requires the development of comprehensive guidelines for educators and AI developers that outline best practices and promote transparency, accountability, and ethical use of AI in education. Ultimately, addressing data privacy and security issues in the context of AI-enabled education is fundamental to ensuring that the benefits of these technologies are fully realized while minimizing the associated risks. 

\subsection{Bias and Discrimination}
The use of AI in education has shown great promise in enhancing teaching and learning experiences. However, the issues of bias \cite{Dwivedi} and discrimination present a significant challenge to the effective and equitable utilization of AI in education. Bias in AI systems is mainly derived from the data they are trained on, which is frequently influenced by historical and societal biases. Large language models such as GPT and Bard are trained on huge amounts of publicly available data from the internet, which contain different points of view. These biases, when embedded in educational AI applications, can lead to a perpetuation of existing disparities, and even exacerbate them, ultimately impacting students’ learning outcomes, opportunities, and access to resources. AI algorithms can inadvertently perpetuate biases present in the data they are trained on, leading to discriminatory \mbox{outcomes  \cite{Ntoutsi}}. Additionally, the use of algorithmic decision making (ADM) by tertiary institutions to admit students to programs raises issues of fairness and objectivity \cite{Marcinkowski}.  It is therefore essential to address these biases and develop fair and equitable AI systems that promote diversity and inclusion in the learning environment \cite{Holstein}. 

Discrimination in AI-based educational tools arises when these systems inadvertently disadvantage specific groups of students based on factors such as race, gender, or socioeconomic background. In content creation, AI-generated material may contain cultural biases, acknowledged by Open AI, in its initial risk analysis and mitigations document for the DALL·E 2 system, where it expresses the following: “DALL·E 2 tends to serve completions that suggest stereotypes, including race and gender stereotypes. For example, the prompt “lawyer” results disproportionately in images of people who are White-passing and male-passing in Western dress, while the prompt “nurse” tends to result in images of people who are female-passing” \cite{Mishkin}. Since a vast majority of the information that exists on the internet has been produced in the Global North, the AI models trained on these data will be molded in the same fashion. Thus, to mitigate the AI divide highlighted in the Oxford Insights Government AI Readiness Index \cite{Oxford2022}, the use of AI in non-Western educational institutions requires further tuning and customization according to the regional and local cultural norms. In assessment evaluation, AI tools might exhibit bias towards certain linguistic or cultural backgrounds, consequently affecting the performance evaluation of students from diverse backgrounds. Additionally, adaptive learning systems, which tailor learning experiences to individual students’ needs, may inadvertently favor students who exhibit learning patterns similar to those of the majority in the training data, thereby (1) widening the achievement gap and (2) potentially exacerbating inequalities. As pointed out in \cite{Borenstein}, the root of the problem is with people, even though there is a need to fix the bias in the data, within the algorithms, and in the outputs. Addressing these issues therefore requires greater collaboration between software engineers, educators, and policymakers to develop transparent, accountable, and inclusive systems to ensure responsible deployment of AI \mbox{in education}.

\subsection{Plagiarism and Academic Integrity}
AI plagiarism has been one of the hottest topics of debate since the introduction of AI chatbots and virtual assistants, such as ChatGPT, Bard, YouChat, Hubspot ChatSpot, Bing Chat, or Vicuna. Whilst chatbots can be responsive pedagogical tools \cite{Chen}, educators have raised the alarm about students using chatbots to compose essays, write computer code, and complete other homework assignments \cite{Sullivan}. The convenience and effectiveness of AI chatbots are tantalizing for students. A survey of 5894 students from across Swedish universities about their use of and attitudes towards AI for learning purposes, both about chatbots (such ChatGPT) and other AI language tools (such as Grammarly), indicated that 95\%  of students were familiar with ChatGPT; 56\% were positive about using chatbots in their studies; 35\%  use ChatGTP regularly; 60\% were opposed to a ban on chatbots, and 77\%  were against a ban on other AI tools (such as Grammarly) in education \cite{Malmstrom}. As online education is growing increasingly popular, an environment where the use of chatbots is particularly easy, the issues of (a) AI plagiarism and (b) detecting academic misconduct with the use of artificial intelligence \cite{Cotton, Kamalov} have become more pressing \cite{Khalil2022, Selwyn}. The new technology is “a clear threat to academic integrity for HEIs, requiring a range of adjustments to be made in both practice and policy” \cite{Perkins}. Thus, it is imperative to identify and implement solutions that address the issue of AI plagiarism. 

Several studies have already considered the effectiveness of Generative AI (ChatGPT, davinci-003) in completing student assignments \cite{AlAfnan, Dwivedi}. Researchers have compared the quality of authentic student work to that of AI. In \cite{Terwiesch}, the researchers considered the performance of ChatGPT (version GPT-3.5) on the final exam of a typical MBA core course, Operations Management. They found that it does an excellent job at basic operations management and process analysis questions, including those that are based on case studies. On the other hand, it makes surprising mistakes in relatively simple math calculations. Another research direction has been the study of anti-plagiarism software in detecting AI work \cite{Wahle}. Most of the currently available anti-plagiarism software is aimed at detecting plagiarism from existing literature \cite{Meo}. Since AI chatbots often produce original content, the off-the-shelves anti-plagiarism software is not well-suited to detect AI plagiarized content  \cite{Khalil2023}. Given the inefficiencies of the traditional software to detect plagiarized content, several attempts have been made to create software that is aimed specifically at identifying AI-generated text. A popular AI plagiarism tool, GPTZero, utilizes perplexity as a measure of the complexity of text, and burstiness---variations of the sentences---to detect AI-generated text. The software creators claim a 95\% detection rate of AI-generated text. Other options include Turnitin Feedback Studio (TFS with Originality, Turnitin Originality, Turnitin Similarity, Simcheck, Originality Check, and Originality Check+), with a 98\% detection confidence rate.

\subsection{Teacher--Student Relationship}
The arrival of AI in education has potential to produce significant transformations in the teacher--student relationship \cite{Rasul}. While AI technology offers promising ability for personalized learning experiences, efficient assessment, and adaptive feedback, it also presents novel challenges to the dynamics of the traditional educational bond \cite{Rudolph}. As the technology continues to permeate the educational sphere, concerns arise regarding the potential dilution of human connection, as well as the erosion of teacher--student rapport, with the consequent impact on the development of socio-emotional skills and a shared sense of classroom community. 

The increased reliance on AI will invariably affect the interpersonal dynamics between teachers and students. The introduction of AI will undoubtedly alter the traditional role of teachers and affect the relationship between the educators and students \cite{Buckingham}. A survey of 1002 K–12 teachers and 1000 students between 12 and 17 in the United States in 2023, commissioned by the Walton Family Foundation, indicated that  51\% of teachers had used ChatGPT (10\% almost every day). Three in ten teachers had used it for lesson planning (30\%), coming up with creative ideas for classes (30\%), and building background knowledge for lessons and classes (27\%); 73\% indicated that they thought the tool can help students learn more.  Interestingly, and in terms of continuing professional development, three-quarters of teachers said using ChatGPT could help them grow as teachers (77\%) \cite{Walton}.   

The consequences of the new teacher--student paradigm can be unexpected. As AI becomes more embedded in content creation and assessment evaluation, one potential outcome could be the decline in teachers’ authority, as students no longer consider them as the ultimate authority. Ensuring that the integration of AI does not undermine the human connection and emotional support provided by educators is essential for the holistic development of students \cite{Lindsey}. It has been argued that AI has potential to engender dramatic and profound changes in teaching and assessment of students, forming an indispensable component of a new paradigm within which we design and deliver education \cite{Yeadon}. The potential issues highlighted in this section require a critical examination of the potential consequences of AI integration in the educational domain, as well as a commitment to the establishment of guidelines and protocols that safeguard the rights and well-being of students. The delicate balancing act between leveraging AI’s potential benefits and addressing the associated challenges underscores the need for a multidisciplinary approach in the ongoing discourse surrounding AI implementation in education. 


\section{Future Directions and Opportunities}
As AI technology continues to advance, it will generate new and unimaginable applications in education. One of the most exciting future opportunities involves the fusion of AI and virtual reality to provide learners with visually rich educational content. Another direction for future application is lifelong learning, where AI is poised to transform the landscape of continuous education and upskilling, laying the foundation for a more adaptable and resilient workforce in the future. On the other hand, as AI permeates multiple facets of daily life, it is important to educate people about AI literacy. Given the power of AI, it is essential to be aware of the ethical consideration when using the technology. A recent example is the launch of the Frontiers of Computing initiative at the University of Southern California, which aim is to embed digital/AI literacies, ethics, and responsibilities across all disciplines. JISC’s AI in tertiary Education: A summary of the current state of play report \cite{JISC} also provides key legal and regulatory frameworks and guidance on the ethical uses of AI.

\subsection{Augmented and Virtual Reality}
Augmented reality (AR), Mixed Reality (MR), and virtual reality (VR, with HoloLens or Meta Quest 2 headsets) are three of the most exciting new technologies which hold a tremendous capacity to enrich human life. In the rapidly evolving landscape of education, the integration of AI with AR, MR, and VR promises to revolutionize learning experiences and pedagogical approaches. AI-driven AR, MR, and VR technologies have the potential to create immersive and interactive educational environments (Metaverse), fostering engagement and enhancing students’ cognitive abilities \cite{Hwang}. These advances enable personalized and adaptive learning pathways, accommodating various learning styles and addressing individual needs. The development of AI-powered AR/VR applications can also provide students with access to simulations and virtual environments, enabling them to acquire practical skills and knowledge in various disciplines \cite{Radianti}. However, the widespread adoption of AI, AR, MR, and VR technologies in education raises numerous ethical and practical concerns that warrant further investigation.

The potential for exacerbating socio-economic disparities and widening the digital divide must be considered, as access to these advanced educational tools may be limited to those who can afford them. In addition, while AI-driven AR and VR applications have demonstrated promising outcomes in various educational contexts, the long-term efficacy and impact on learning outcomes remain to be thoroughly assessed. As a result, future research should focus on addressing these challenges, ensuring the responsible and equitable integration of AI-driven AR and VR technologies into educational ecosystems.

\subsection{Lifelong Learning and Skill Development}
In today’s fast-changing job landscape, lifelong learning and skill development have emerged as critical aspects of modern education. The popularity of online education platforms such as Coursera, Edx, and Udemy, which allow individuals to upskill \cite{Santandreu2019} or even transition to completely new fields of expertise, shows the importance of continued education beyond college. We anticipate the role of AI in this domain to be transformative. AI holds the potential to have a big impact on the way individuals engage with learning processes throughout their lives, enabling highly personalized educational experiences that adapt to the evolving needs and abilities of learners. Specifically, AI-driven learning platforms could seamlessly integrate formal and informal learning opportunities, thereby encouraging self-directed skill acquisition and continuous intellectual growth. 

We expect AI to address the challenges inherent to lifelong learning and skill development by delivering a dynamic and customizable educational framework. It can play a pivotal role particularly in the context of the rapidly evolving job market and the need for continuous upskilling \cite{Bessen2019}. AI-driven learning platforms can help individuals identify skill gaps, access relevant learning resources, and track their progress, facilitating career growth and adaptability \cite{Kabudi}. One of the key aspects of AI is the potential to harness massive datasets to identify emerging trends, skill gaps, and workforce requirements. The extracted information can be used to design and deliver education and training programs that align with the evolving demands of the global economy. By intelligently identifying and addressing individual skill gaps, AI systems can facilitate efficient upskilling and reskilling, enabling individuals to adapt more effectively to shifting job markets and technological advancements.

\subsection{AI Literacy and Ethics Education}
As the proliferation of AI-based systems and applications continues to accelerate, the need for individuals to possess a foundational understanding of AI principles, techniques, and potential applications becomes increasingly crucial. AI literacy will be essential not only for engineers but for all professions. Educators must prioritize the integration of AI literacy into curricula, ensuring that students are equipped with the knowledge and skills necessary to navigate the rapidly evolving AI landscape. This focus will enable students to better comprehend and utilize AI technologies as well as prepare them for careers in the AI-dominated future job market \cite{Borenstein}. The increased integration of AI into various aspects of society requires that we equip students with AI literacy and ethical \mbox{awareness \cite{Long}}. Incorporating AI and ethics education in curricula can help students understand the implications and responsibilities associated with AI technologies, fostering responsible innovation and informed decision making  \cite{Boddington}. 

As the impact of AI becomes more pervasive, ethical considerations surrounding its deployment and use will become paramount \cite{Borenstein}.  Initial findings of an ongoing survey of more than 450 students in Hong Kong and pilot focus group panels with 13 Australian students indicated that educating students about risks, biases, and limitations was \mbox{crucial \cite{Liu2023b}}. Consequently, ethics education must be incorporated into AI-related curricula. It is crucial to instill in students an awareness of the ethical implications and potential risks associated with AI technologies. By fostering a critical mindset that balances the benefits of AI with its potential harms, educators can promote responsible AI usage to mitigate adverse societal consequences. The proposed approach will support the development of a generation of AI-literate and ethically conscious individuals who are equipped to confront and navigate the complex challenges presented by the rise of AI.


\section{Conclusions}
{
In this paper, we reviewed and analyzed the current literature to better understand the potential effects of AI in education. We aimed to provide both a general overview as well as a more specific discussion of various aspects of the subject. Our review focused on three major themes: applications, benefits, and challenges.
We found that the advent of AI brings tantalizing possibilities and applications in the education sector. Its impact is multifaceted and holds the potential to revolutionize the way learning is delivered and experienced. 
As we enter the new era in education, the present study allows for a moment of reflection based on the aggregate survey of the existing knowledge.
}

The applications of AI in education include personalized learning, intelligent tutoring systems, assessment automation, and teacher--student collaboration, which can help improve learning outcomes, efficiency, and global access to quality education. The scalability of AI means that its benefits can be shared by large swaths of the society, providing high quality education around the world. 
While AI has the capacity to make a significant positive impact on education, it is important to keep in mind the dangers of misusing AI. There are several concerns related to the deployment of AI; these include data privacy, security, bias, and teacher--student relationships, and they must be addressed to ensure the responsible and ethical implementation of AI in education. To meet the challenges presented by the rise of the technology, AI literacy and ethics education must become a part of the curricula. By leveraging these advancements, educators and policymakers can work towards creating inclusive, equitable, and effective learning environments that cater to the diverse needs of learners in the 21st century.

While this study presented a theoretical overview of the potential effects of AI in education and can serve as a springboard for the development of the subject, an empirical study is required to provide more concrete results. In the future, studies based on student cohorts measuring the difference in the learning outcomes between AI-driven and traditional teaching methods or teacher surveys measuring the actual number of saved hours when using automated grading systems are needed.

\end{document}